# A Memristor-Inspired Computation for Epileptiform Signals in Spheroids

Iván Díez de los Ríos[1], John Wesley Ephraim[2], Gemma Palazzolo[2], Teresa Serrano-Gotarredona[1], Gabriella Panuccio[2], Bernabé Linares-Barranco[1].

[1] Instituto de Microelectrónica de Sevilla (IMSE-CNM), CSIC and Univ. de Sevilla, Sevilla, SPAIN
[2] Istituto Italiano di Tecnologia (IIT), Genoa, Italy

*Abstract*—In this paper we present a memristor-inspired computational method for obtaining a type of running "spectrogram" or "fingerprint" of epileptiform activity generated by rodent hippocampal spheroids. It can be used to compute on the fly and with low computational cost an alert-level signal for epileptiform events onset. Here, we describe the computational method behind this "fingerprint" technique and illustrate it using epileptiform events recorded from hippocampal spheroids using a microelectrode array system.

## I. INTRODUCTION

Regenerative medicine is a promising branch of health science that aims at restoring brain function by rebuilding brain tissue. However, repairing the brain is one of the hardest challenges and we are still unable to effectively rebuild brain matter. Epilepsy is particularly challenging due to its dynamic nature caused by the relentless brain damage and aberrant rearrangements of brain rewiring. To overcome the biological uncertainty of canonical regenerative approaches, innovative solutions are being proposed based on intelligent biohybrids, made by the symbiotic integration of bioengineered brain tissue, neuromorphic microelectronics and artificial intelligence, to effectively drive self-repair of dysfunctional brain circuits. The EU project HERMES [i] fosters the emergence of a novel biomedical paradigm, rooted in the use of biohybrid neuronics (neural electronics), which we name enhanced regenerative medicine. In this project one of the components is an implantable "Neuromorphic Computing System" (NCS), based on memristors, that can establish a bi-directional communication with a graft to be implanted in an epileptic brain, replacing damaged epileptic tissue. This concept is illustrated in Fig. 1. The NCS interacts with the graft to guide its integration within the host brain. As the latter might still generate seizures during the regeneration process, the NCS learns a cost function to predict the probability of seizure build-up and develops a stimulation policy to halt or prevent it. Since the NCS has to be a low power implantable system, during its training phase, it is supervised by an external Artificial Intelligence system (AI) running on high power servers. Once the NCS has learned its cost function, the AI system is disconnected. With this long-term vision in mind, as a first step, an initial goal is to develop an NCS capable of interacting with the graft, predicting ictal activity and establishing an adaptive stimulation policy to counteract it. In order to recognize when seizure generation probability increases, while relying on low computational costs, we propose here a simple low-computational frequency decomposition memristor-operation-inspired computing function, which we call the "memristor-transform". By extracting a simple tunable and adaptive cost-function from this "memristor-transform", one can then further develop a stimulation policy to halt the seizure or prevent its onset. This memristor-inspired computation can be readily implemented in highly compact low-power hardware. In this paper we present the concept of this "memristor-transform" and illustrate it with in vitro microelectrode array (MEA) recordings of rodent hippocampal spheroids.

The rest of the paper is organized as follows. Section II describes the in vitro measurement setup, Section III explains the computational "memristor-transform", and Section IV illustrates how to apply this concept to four filtered versions of spheroid MEA recordings to obtain a "fingerprint" that can be used to compute an alert-level signal for an upcoming ictal event.

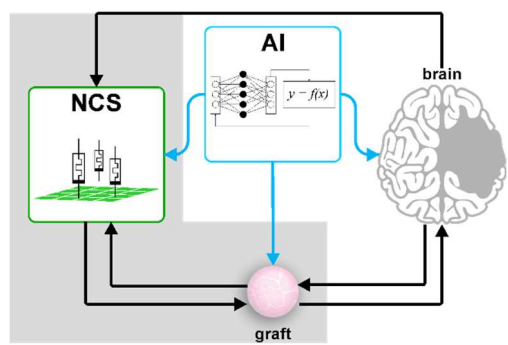

Fig. 1.Enhanced regenerative medicine based on biohybrid constructs. The neuromorphic computing system (NCS) guides the integration of the graft within the host brain. Artificial Intelligence (AI) supervises the NCS and guides its learning.

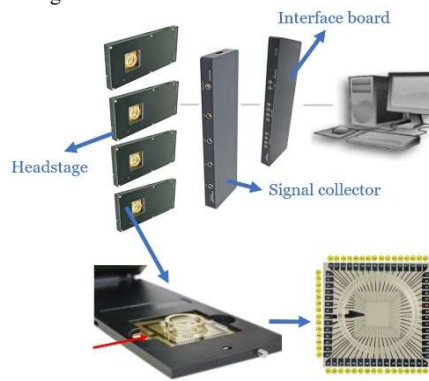

Fig. 2.Microelectrode array set-up.

## II. MEA SYSTEM DESCRIPTION

There is a vast literature body on biomedical systems for EEG-based seizure detection and prediction [1]-[19], although normally their computations are not memristor-inspired. Here we focus on direct MEA based in vitro measurements, and using a memristor-inspired computation. Fig. 2 illustrates a MEA2100-mini-HS60 system for in vitro recordings[ii]. The MEA headstages are connected to a signal collector unit, plugged onto an interface board, which connects the system to a host PC. In our setup we used only one headstage. All the components are from Multichannel Systems, Germany.

Individual hippocampal spheroids were taken from the incubator immediately prior to the recording and accommodated the MEA and let habituate for 20 minutes.

This work was funded by EU H2020 grant 824164 "HERMES".

Recordings were performed at ~37°C, achieved with the use of a custom-made heating lid covering the headstage along with the warming of the MEA amplifier base (temperature set at 37°C). The recording medium was equilibrated at pH~7.4 through humidified carbogen delivered via a tubing connected to the heating lid and consisted of (mM): NaCl 117, KCl 3.75, KH2PO4, 1.25, MgSO4 0.5, CaCl2 2.5, D-glucose 25, NaHCO3 26, L-Ascorbic Acid 1. Signals were acquired via an 8x8 3D MEA (TiN electrodes, diameter 12 μm, height 80 μm, inter-electrode distance 200 μm, impedance ~150 kΩ, internal reference electrode), sampled at 20 kHz and low-pass filtered at 10 kHz before digitization. Each recording session lasted 20 minutes. For off-line analyses, signals were first averaged to an equivalent sampling frequency of 400 Hz.

### III. MEMRISTOR TRANSFORM

In a recent paper, Liu et al. [20] presented a method based on physical memristors for discerning between pre-ictal and interictal recordings. For this, they used a public open-access benchmark, the Kaggle Seizure Prediction Dataset [21]. This dataset includes 16-channel 400Hz sampling-rate iEEG recordings from 5 different dogs. These belong to two categories:

a) **Interictal**, recorded at least one week before or after a seizure

b) **Pre-ictal,** recorded in the range of one hour before a seizure

The authors selected 2.4-second clips which, sampled at 400 Hz (2.5 ms), contain 960 data points on each of the 16 channels. These 960 data points are grouped into 64 separate data segments of 15 data points. This way, for each 2.4s clip, there is a total of 16x64=1024 15-data-point segments. Each of these 1024 segments is applied to a physical memristor to erase it [22]-[23], and a "*finger-print*" (or *feature*) matrix is built with the conductance change $\Delta G$ the memristors had experimented. This fingerprint from physical memristors is then used to train a classifier to successfully distinguish between Interictal and Pre-ictal recordings. In the next subsection we disclose the ideal mathematical computation the memristors perform to obtain such fingerprints. The underlying computation is fairly simple and inspired the "memristor-transform" explained in Section IV. Example 1k-matrices are shown in Fig. 3 for 3 different interictal and pre-ictal 2.4s clips, using the ideal underlying computations.

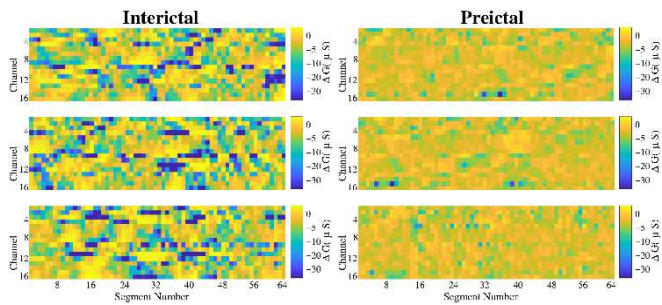

Fig. 3. Illustration of "finger-print" matrices, each for one specific 2.4s 16-channel clip. Half of the finger-prints correspond to interictal recordings, and the other half to preictal ones.

#### A. Modeling the memristor computations

Liu et al. exploited the memristor erase operation to compute the aforementioned finger-prints. For this, all memristors were initially set at high conductance state ("ON" state) and then, by using amplitude dependent erase pulses, they progressively decreased the conductance for each memristor. Let us assume that the decrease in conductance can be modelled as [24]

$$\Delta G = \propto(x) G^{\gamma(x)} \quad (1)$$

where $x$ is the applied erase amplitude, and coefficients $\alpha(x)$ and $\gamma(x)$ are fitted for each curve. It turns out that a fairly good fitting is obtained when $\alpha(x)$ and $\gamma(x)$ have an amplitude dependence given by

$$\begin{aligned}\gamma(x) &= -Ax + B \\ \propto(x) &= -Ke^{Px}\end{aligned} \quad (2)$$

where $A$, $B$, $K$, and $P$ are fitting parameters. The values we obtained for these parameters were $A = 7V^{-1}$, $B = 16.1$, $K = 6.31 \times 10^{-30}$, $P = 32.24 V^{-1}$. This way, the "memristor-transform" consists of applying eqs. (1-2) to $n$ consecutive samples of a recorded signal, or, as we will see in Section IV, to $n$ consecutive samples of filtered versions of a recorded signal. Fig. 4 shows the conductance decrease steps when applying 15 consecutive constant-amplitude erase pulses. The erase pulse amplitude changes between 1.3 V and 1.8 V in steps of 0.1 V.

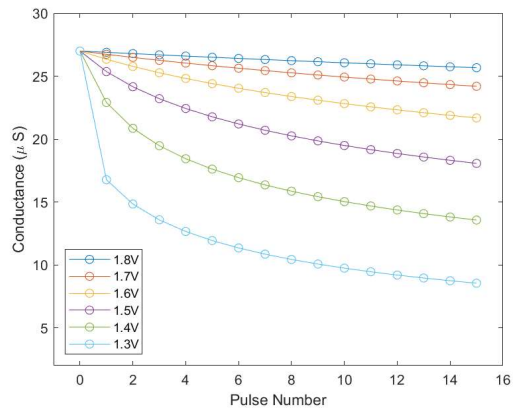

Fig. 4. Conductance reduction obtained by progressively applying 15 erase pulses of constant amplitude.

### IV. PROCESSING OF SPHEROID RECORDINGS WITH THE MEMRISTOR-TRANSFORM

Fig. 5(a) shows a 20-minute recording from a spheroid generating a mixed pattern of seven ictal events together with interictal events. Fig. 5(b) shows a 90 second zoom-in focusing on the 5th ictal event. Fig. 5(c) illustrates a 7 second zoom-in, and Fig. 5(d) a 300 ms zoom-in. Each zoom-in is approximately one order of magnitude lower, illustrating the richness of mixture of slower to faster signal components. As a first-order quick analysis, we would like to extract the different timing components for each order of magnitude. To do this, we compute running averages at different time-scales:

i. A type of reference signal which computes a DC average over the 10 s previous samples. Let us call it **f0** component.

ii. A slow variation signal which computes the average over the 1 s previous samples. Let us call it **f1**. This

average signal should be capable of capturing the slowly varying changes of the main events.
iii. A faster variation signal which computes the average over the 100 ms previous samples. Let us call it **f2**. This one can capture faster variations.
iv. And a fastest one, computing the average over the 10 ms previous samples. Let us call it **f3**.
v. Additionally, we consider also the original 400 Hz sampled signal for completeness, which we name **f4**.

These five signals {**f0, f1, f2, f3, f4**}, can be considered low-pass filtered versions of the original **f4** signal. The different time-range components can be obtained by computing the differences among them. For example, timing characteristics in the *1s-to-10s* range can be extracted by computing **f1-f0**; timing characteristics in the *100ms-to-1s* range can be extracted by computing **f2-f1**; timing characteristics in the *10ms-to-100ms* range can be extracted by computing **f3-f2**; and the fasted timing characteristics can be extracted from **f4-f3**. These signals can be interpreted as bandpass versions, or as computing the changes of one time-scale average with respect to the previous slower time-scale average (like a derivative).

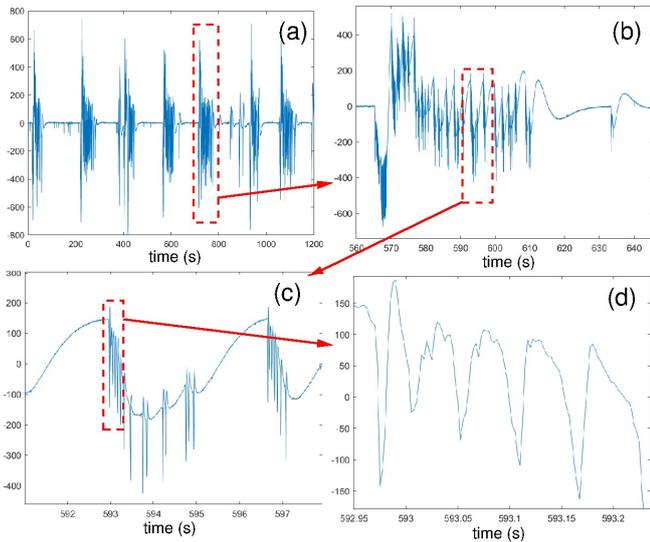

Fig. 5. (a) 20-minute (1200 s) recording of a spheroid sample, including mixed interictal and ictal events. (b) 90 s zoom-in view of 5th event in (a). (c) 7 s zoom-in view of (b) focusing on a segment that shows slower variations mixed with faster ones. (d) 300ms zoom-in of (d) focusing on the higher fequency variations.

Fig. 6 illustrates this. There are three panels, each for different time scales: Fig. 6 (a) shows a 60 s view, Fig. 6 (b) shows an 8 s view, and Fig. 6 (c) shows a 500 ms view. In each panel, the top subplot (x1) shows the original signal **f4** (down sampled at 400 Hz), the center subplot (x2) shows "low-passed" signals {**f3, f2, f1, f0**}, and the bottom subplot (x3) shows the "band-pass" signals {**f43, f32, f21, f10**}.
In Fig. 6(a) the 60 s view shows the fifth event in Fig. 5(a). Here we can see that signal **f0** is filtering out most of the faster variations and is like a running reference level. Signal **f1** follows closely the slow-varying 1sec-range ups-and-downs while it filters out the faster variations. Therefore, **f10** captures the slow 1s-range changes in **f1** with respect to **f0**. On the other hand, **f21** captures faster changes, as it follows the **f2** (averaged over previous100ms) changes with respect to **f1** (averaged over previous 1s) changes. Fig. 6 (b) shows a 7s view. Here we can see that **f0** is practically constant. **f1** follows (with some delay) a clean averaged version of the original signal, filtering out all faster transitions. **f2** flows more closely the original signal but filtering out the very fast transitions, which are still present in **f3**. Consequently, we can see that **f21** highlights changes in the second to 100ms range, while **f32** captures changes that are faster than seconds, in the 100ms ranges. Finally, Fig. 6 (c) shows a 500ms view, where we can see that signals **f0**, **f1**, and **f2** are fairly constants, while **f3** is able to follow the very fast changes in the original signal. This way, **f32** tracks the fast variations in the original signal, while **f43** resembles its derivative.

Fig. 7 shows a different 200 second section of the recording in Fig. 5 (a), but now adding in the bottom subplot the finger-prints of the memristor-transform illustrated in Fig. 3 applied to the four timing components {**f10, f21, f32, f43**}. These fingerprints highlight the amplitudes of the different speed/frequency components. One can see that during one cycle (blue followed by yellow in Fig. 7) of the slower speed component **f10**, there appear more than one cycle of the higher speed components **f32** and **f43**. Additionally, low speed component cycles in **f10** show rather short separations and durations. This contrasts strongly with the interictal signals, like the one illustrated in Fig. 8. Here one can see that each event has just one cycle for components **f10**, **f21 f32**, and **f43**. Also, one can see that event separations are typically in the range of 10-20 seconds or more (except the first two). This suggests that one could use the event sizes and separations of the lower speed components **f10** to evaluate how spheroids are transitioning from interictal to ictal regimes [25]. An additional hint seems to be the number of faster component cycles **f21-f43** that might appear during one low-speed cycle, spread over its duration. This way, one could compute a "cost function" or "alert-level signal" that depends on event properties, like frequency components widths and separation, as well as number of faster cycles within a low speed event (see red dashed line in center subplot of Fig. 7 and Fig. 8). This alert signal could then be used to establish a stimulation policy to halt or prevent ictal events while minimizing the number of delivered pulses.

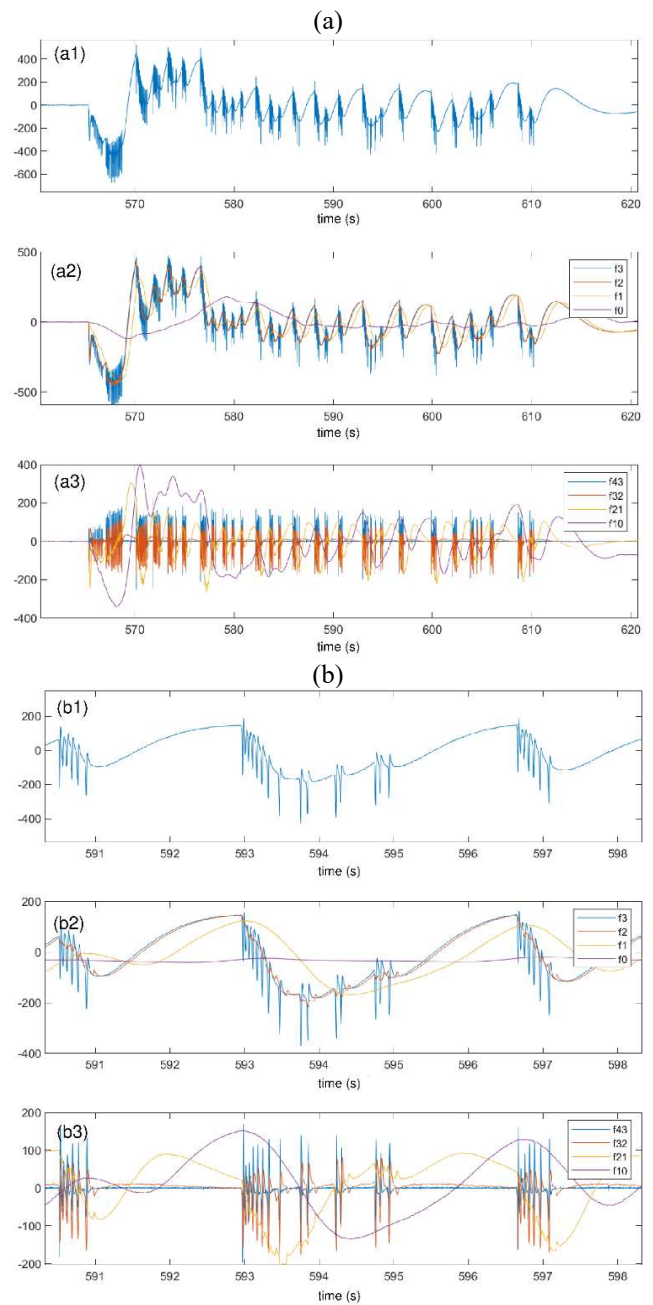

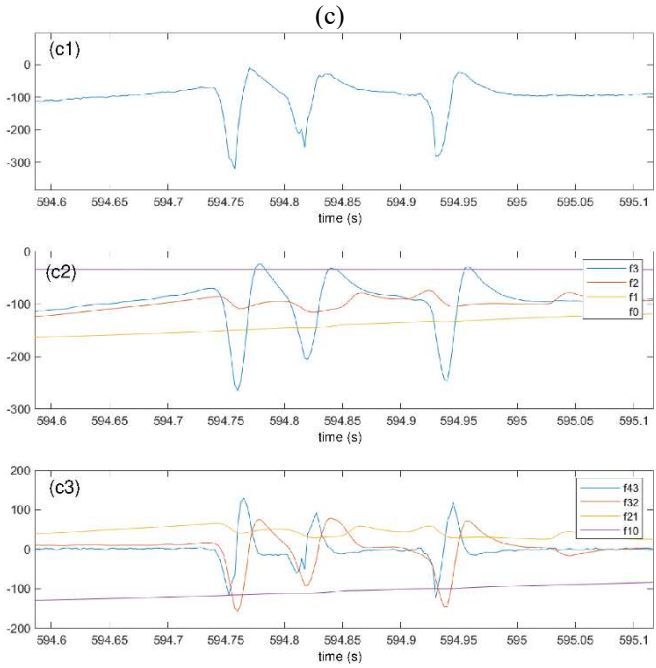

Fig. 6. Illustration of timing properties extraction for the signal in Fig. 5(b). There are three panels (a-b). Each panel shows a different time scale, and each includes three subplots 1-3. Subplots (x1) are the original f4 signals sampled at 400Hz rate. Subplots (x2) show signals f3, f2, f1, and f0. Subplots (x3) show f43, f32, f21, and f10. Panel (a) shows 60 second view. Panel (b) shows an 8 second view. Panel (c) shows a 500ms view.

## V. CONCLUSIONS AND FUTURE WORK

We have proposed a computationally efficient "memristor-transform" to obtain a frequency dependent fingerprint of epileptiform signals. This easy-to-compute and informative transform could eventually help in estimating probabilities of ictal events onset. This fingerprint resembles a type of spectrogram, but with lower computational cost. From this fingerprint, a continuous alert-level signal could be derived, also with minimum computational cost, for computing a probability of an impending ictal event. Future work includes experimenting this technique in a closed-loop in vitro setup, and devise stimulation policies for interacting with the spheroids in real-time and learn how to prevent ictal discharges, while minimizing the number of stimulation pulses.

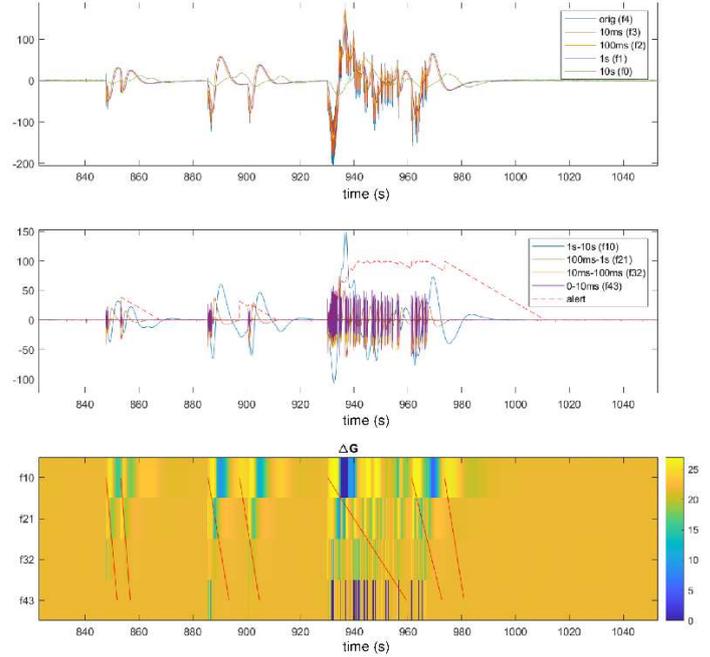

Fig. 7. Same recording as in Fig. 6, but now adding in the bottom subplot the memristor-based computational finger-print. Red oblique lines in bottom graph indicate events. Dashed line in center graph is a computed "alert" signal. "Alert" signal increases with short inter-event separations and high frequency components, while it decays with time by default.

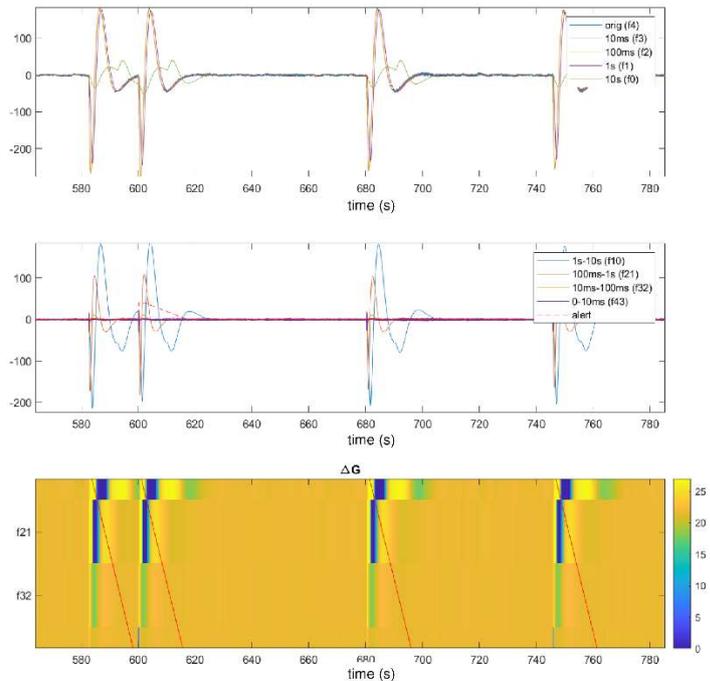

Fig. 8. Young spheroid signal showing interictal-like activity. Alert signal only rises a bit at second event, as it is very close to the first one.

---

[i] Hybrid Enhanced Regenerative Medicine Systems. EU H2020 FET project. https://hermes-fet.eu
[ii] https://www.multichannelsystems.com/products/mea2100-systems